\documentstyle{article}
\begin{document}
\title{Stochastic Schr\"odinger equation from an interaction
with the environment}
\author{Z.Haba\\Institute of Theoretical Physics, 
\\ University of Wroclaw,
Wroclaw,Poland}
\date{}
\maketitle
\begin {abstract}
We consider a class of models describing a quantum oscillator in
 interaction with an environment. We show that models of continuous
 spontaneous localization based on a
stochastic Schr\"odinger equation can be derived as an approximation
to purely deterministic Hamiltonian systems. We show an exponential
decay of off-diagonal matrix elements in the energy representation.
\end{abstract}

\section{Introduction}

       The conventional quantum mechanics assumes
 a sharp division of the world into microworld of the system and
 macroworld of the measuring devices.Such a theory can be considered 
 as a theory
 of an ideal measurement. However, in the contemporary mesoscopic physics
 such a division may be not adequate. We can imagine a direct observation
 of a mesoscopic body interacting with an invisible microparticle system.
 Such a record cannot be considered as  a
 measurement in the conventional quantum mechanics (e.g. the behavior
 of the mesoscopic body can be reversible). Nevertheless, it gives us
 some information about the microsystem. Such a situation inspires
 a study of the influence of a large (possibly
 infinite) system on the behavior of a single particle. It seems that
 together with an investigation of a classical limit of the quantum mechanics
 done so extensively on the level of the wave function a study of the
 classical limit of the quntum theory of measurement is needed as well.
 It is well-known that we encounter difficulties with an interpretation
 of quantum mechanics if it is applied to macroscopic bodies. However,
 it may be that the troublesome interference principle has no
 observational consequences when applied to realistic macroscopic bodies
 ( apart from interesting situations on the verge of quantum and classical
 worlds \cite{Legg}). A research in this field ( which could in general be
 called a problem of decoherence) evoked interest recently \cite{Zur1}
 \cite{Gell}\cite{Joos}.The problem of decoherence seems to be closely
 related \cite{Digi}to another old problem of quantum measurement theory 
 that of the
 reduction of the wave packet. There appeared recently a conceptually simple
 theory of the wave packet reduction \cite{Ghi2}\cite{Gisi}.
 In this theory the wave packet reduction is achieved through an addition
 of noise to the Schr\"odinger equation.Such
 a theory could be considered either as a phenomenological description
 of experiments \cite{Joos} or as a fundamental modification of
 quantum mechanics which is supposed to describe both microworld and
  macroworld \cite{Ghir}\cite{Ghi2}.

       In this paper we do not intend to enter the basic epistemological
  problems. We investigate some simple models of an interaction of an
  oscillator with a reservoir ( the simplest one is a text-book example
;see refs. \cite{Loui} \cite{Hake}).The model is usually considered
as a typical application of the quantum Langevin equation.
 We have shown in our earlier papers \cite{hab1} \cite{hab2}
 that quantum dynamics can be expressed in terms of conventional diffusion
 processes (time-ordering replaces the non-commutativity). Such a stochastic
 description can be considered as a mathematical version of the Feynman
 path integral \cite{hab3}. We think that
 such an approach is especially useful for a description of the dynamics
 of a subsystem.We can investigate
  the dynamics of a subsystem
 through an exact elimination of the coordinates of the environment (
 such a description is well-known for classical stochastic systems
 \cite{Zwan}). In principle, this is the same strategy as suggested first
 by Feynman and Vernon \cite{Feyn} and applied e.g. in refs.\cite{Legg}
 \cite{Zur1}. However,instead of using the path integral we eliminate
 the coordinates of the environment on the level of stochastic equations.
  We get (without any approximation) closed equations
   for a particle moving in a reservoir. Our approach is based on the
 classical probability applied as a technical tool to standard quantum 
 mechanics.
 It comes out from our study that a noise which initially enters the 
 calculations
 as a technical tool acquires properties and functions of a physical noise.
 The Brownian motion which initially is a realization of the sum over
 paths describing via the Feynman formula the unitary Schr\"odinger evolution
 acquires an interpretation of a dissipation when we pass to the subsystem.
 There is still another source of noise (which we call a deterministic noise).
 The reservoir degrees of freedom
 evolving according to
 the deterministic Schr\"odinger dynamics in the limit of an infinite
 number of degrees of freedom model quite well the white noise. This model
 of noise is  similar to the mechanical model of the Brownian motion
 discussed in the well-known paper of Ford et al \cite{Kac}.
We show that the Brownian noise together with the deterministic noise  give
 in a proper limit
the stochastic Schr\"odinger equation. As a consequence of the stochastic
Schr\"odinger equation we obtain a model of a continuous
spontaneous localization \cite{Ghi2} . We show that in the energy
representation (for a sufficiently large energy) the density matrix decays
exponentially to its diagonal (decoherence). A localization in
momentum  ( or space) is shown to hold true only in a limit of small
(resp. large ) oscillator frequency.

 \section{Stochastic representation of quantum dynamics}

In our earlier papers [14]-[16] we have described the unitary Hamiltonian
time evolution of pure states in terms of the Brownian motion. This
formalism can be considered as a mathematical version of the Feynman
path-integral. On a formal level the method consists in the
replacement of the integral over real
paths $q(.)$ by an integral over complex paths $\sqrt{i}q (.)$ (where
$i$ is the imaginary unit). As a next step we change variables in the
functional integral. The change of variables is expressed as a solution
of a stochastic differential equation (depending on the Brownian motion).
Then, the non-Gaussian functional integral can be reduced to the Gaussian
integral over solutions of the stochastic equation.The aim of this formalism
 is to express in an explicit form
the solution of the Schr\"odinger equation. Hence, it is completely
equivalent to standard quantum mechanics.

Let $U_{t}$ (t$\geq0$)be a unitary Schr\"odinger evolution determined by the
Hamiltonian
\begin{equation}H=-\hbar^{2}\frac{1}{2m}\triangle + V
\end{equation}
Assume we know
 \begin{math}
| \chi>_{t}=U_{t}| \chi>
\end{math}
 ( $\chi(x)$ will denote $|\chi>$ in the coordinate
representation, where $x\in R^{n}$).
  Let the initial condition $\psi$ for the Schr\"odinger equation be of 
  the form
$\psi(x)=\chi(x)\phi(x)$ where $\phi$ is an analytic
function.Then, the solution $\psi_{t}$ of the Schr\"odinger equation with the
initial condition $\psi$ can be expressed as $\chi_{t}\phi_{t}$,
 where
$\phi_{t}$ is the solution of the equation
\begin{equation}
\partial_{t}\phi_{t}=\frac{i\hbar}{2m}\triangle\phi_{t}
+\frac{i\hbar}{m} \chi_{t}^{-1}\nabla\chi_{t}\nabla\phi_{t}
\end{equation}
with the initial condition $\phi$.

    We express the solution of eq.(2) by a stochastic process \cite{hab1}( for
time-dependent diffusions see ref.\cite{Frei}). Then, the unitary Schr\"odinger
evolution generated by the Hamiltonian (1) can be expressed in the form
\begin{equation} (U_{t}\psi)(x)=\chi_{t}(x)E[\phi(q_{t}(x)]
\end{equation}
where $q_{t}(x)$ is a complex diffusion process starting at t=0 from x
and solving the stochastic differential equation (here $0\leq \tau \leq
t $)
\begin{equation} dq_{\tau}=\frac{i\hbar}{m}
 \chi_{t-\tau}^{-1} \nabla \chi_{t-\tau}(q_{\tau}) d\tau
 + \lambda \sigma db_{\tau}
\end {equation}
where
\begin{displaymath}
\lambda=\frac{1}{\sqrt{2}}(1+i)
\end{displaymath}
and
\begin{displaymath}\sigma=\sqrt{\frac{\hbar}{m}}
\end{displaymath}
The Brownian motion $b_{t}$ is defined as the Gaussian process with
independent increments and the variance
\begin{displaymath} E[b_{t}^{2}]=t
\end{displaymath}
In order to express the solution of the Schr\"odinger equation for 
negative time
we may apply the complex conjugation
\begin{displaymath}
\overline{U_{t}\overline{\psi}}=\psi_{-t}
\end{displaymath}

The action of arbitrary operators on states as well as correlation
functions in arbitrary states can now be expressed as expectation values
with respect to the Brownian motion \cite{hab1}.For example
\begin{equation}
<\chi|F(x_{t})G(x)|\chi>=<\chi|U_{t}^{+}F(x)U_{t}G(x)|\chi>
=E[\int dx |\chi_{t}(x)|^{2}F(x)G(q_{t}(x))]
\end{equation}
When $F=1$ then eq.(5) expresses a generalization of the notion of an invariant
measure to time-inhomogenous processes ( if $\chi$ is a stationary state
then $|\chi(x)|^{2}$ is exactly the invariant measure for $q_{t}$ \cite{Frei})
 i.e.
\begin{displaymath}
E[\int dx |\chi_{t}(x)|^{2}G(q_{t}(x))]=\int dx |\chi(x)|^{2} G(x)
\end{displaymath}

     Let us consider as an example a model of independent oscillators
  with $m=1$ (here and later we do not specify the number of oscillators,it can
  be infinite)
 \begin{equation} H_{0}=-\frac{\hbar^{2}}{2}\sum_{k}
  \frac{\partial^{2}}{\partial x_{k}^{2} } + \sum_{k}\frac{1}{2}
  \omega_{k}^ {2}x_{k}^{2}
  \end{equation}
 The ground state solution of the Schr\"odinger equation
with the Hamiltonian (6) reads
 \begin{equation} \chi_{0}(x)=exp(-\frac{1}{2\hbar}\sum_{k}\omega_{k}
 x_{k}^{2})
 \end{equation}
 The stochastic equation (4) takes the form
 \begin{equation} dq_{k}=-i\omega_{k}q_{k}dt + \lambda\sigma db_{k}
 \end{equation}

The solution of eq.(8) with the initial condition x reads
\begin{equation}
q_{k}(s)=exp(-i\omega_{k}s)x_{k}+
\lambda\sigma\int_{0}^{s}exp(-i\omega_{k}(s-\tau))db_{k}(\tau)
\end{equation}
\section{An interaction with the environment}
The formalism of sec.2 can easily be applied to Hamiltonians with explicitly
known $\chi_{t}$.An example of such a class of models
are Hamiltonians ( often encountered in
quantum optics) of the form
\begin{equation}
H_{S}=\sum_{\mu \nu}h_{\mu\nu}(A,A^{+})A_{\mu}^{+}A_{\nu}
\end{equation}
where $A_{\mu}$ is the anihilation operator,$A_{\mu}^{+}$ the creation operator
and the Greek index indicates the space coordinate. In the case of the
Hamiltonian (10) the harmonic oscillator$^{\prime}$s ground state
(7) is also the ground state of $H_{S}$.Hence,we may take $\chi_{0}$ as
$\chi_{t}$ in eq.(3).

We add to $H_{S}$ the coupling to the reservoir. In this paper we restrict
ourselves to the simplest one-dimensional model corresponding to $h=1$.
In such a case
we obtain a standard model of an interaction with the environment (see
\cite{Loui},\cite{Hake} and \cite{Ford} for a recent review).
  \begin{equation}H=\hbar\omega_{0}A^{+}A + \sum_{k}\hbar\omega_{k}
 a_{k}^{+}a_{k} + \sum_{k}g_{k}A^{+}a_{k}+\overline{g_{k}}a_{k}^{+}A
 \end{equation}
 Using the Schr\"odinger representation of the creation and anihilation
 operators
 \begin{displaymath}a=\sqrt{\frac{\hbar}{2 \omega}}\frac{\partial}
 {\partial x}+\sqrt{\frac{\omega}{2\hbar}} x
 \end{displaymath}
 we express H in the form (1). It is clear that the product of oscillator
 ground states is the ground state for H.In order to simplify the
 discussion we assume
 \newpage
 that
 \begin{displaymath}
g_{k}=iv_{k}
 \end{displaymath}
 are purely imaginary ( if $g_{k}$ had a real part this would change only the 
 form of the 
 noise in the
 stochastic equation (4)). Then,taking as $\chi_{t}$ in eq.(4) the ground state
(of $Q$ and $q_{k}$ oscillators) we obtain the following equations
 \begin{equation}
 dq_{k}=-i\omega_{k}q_{k}dt+v_{k}\sqrt{\frac{\omega_{0}}
 {\omega_{k}}}Qdt+\lambda \sigma db_{k}
 \end{equation}
 \begin{equation}
 dQ=-i\omega_{0}Qdt-\sum_{k}v_{k}\sqrt{\frac{\omega_{k}}{\omega_{0}}}q_{k}dt
 +\lambda\sigma db
 \end{equation}

We solve first eq.(12) for $ q_{k}$. We obtain
\begin{equation}
\begin{array}{l}
q_{k}(t)=exp(-i\omega_{k}t)x_{k}+v_{k}\sqrt{\frac{\omega_{0}}{\omega_{k}}}
\int_{0}^{t}exp(-i\omega_{k}(t-s))Q(s)ds +
\cr
+\lambda\sigma\int_{0}^{t}exp(-i\omega_{k}(t-s))db_{k}(s)
\end{array}
\end{equation}
Let us introduce the notation
\begin{equation}
N_{D}(s)=\sum_{k}v_{k}x_{k}\sqrt{\frac{\omega_{k}}
{\omega_{0}}}exp(-i\omega_{k}s)
\end{equation}
and
\begin{equation}
N_{R}(s)=\lambda\sigma \sum_{k}v_{k}\sqrt{\frac{\omega_{k}}{\omega_{0}}}
\int_{0}^{s}exp(-i\omega_{k}(s-\tau))db_{k}(\tau)
\end{equation}
corresponding to a "deterministic" and random noise.
We are interested in a computation of correlation functions in eq.(5) with
 $\chi_{t}$ defined in eq.(7).
Then,in such computations $x_{k}$
play the role of independent Gaussian random variables with the covariance
\begin{equation}
E[x_{k}x_{r}]=\frac{\hbar}{2 \omega_{k}}\delta_{kr}
\end{equation}
Hence,the noise $N_{D}$ is a Gaussian complex stochastic process with the
covariance
\begin{equation}
E[N_{D}(s)N_{D}(\tau)]=\frac{\hbar}{2 \omega_{0}}\sum_{k}v_{k}^{2}
exp(-i\omega_{k}(\tau+s))
\end{equation}
\begin{displaymath}
E[\overline{N_{D}(s)}N_{D}(\tau)]=\frac{\hbar}{2\omega_{0}}
\sum_{k}v_{k}^{2}exp(-i
\omega_{k}(\tau-s))
\end{displaymath}
This noise is of the same order as the $ N_{R} $ noise (16).
 Next,let us denote
\begin{equation}
K(s)=\sum_{k}v_{k}^{2}exp(-i\omega_{k}s)
\end{equation}
We insert the solution $q_{t}$ of eq.(14) into eq.(13) for Q. We 
obtain an equation
describing an oscillator moving in a reservoir
\begin{equation}
dQ=-i\omega_{0}Qd\tau-\int_{0}^{\tau}K(\tau-s)Q(s)ds d\tau
- N_{D}(\tau)d\tau - N_{R}(\tau)d\tau
+\lambda\sigma db
\end{equation}
It can be seen from eq.(19) that K(s) is positive for small s, hence it
fulfills the role of a dissipation.$N_{D}$,$N_{R}$ and b add to a random
external force.

\section{Stochastic Schr\"odinger equation from an interaction with the
environment}
In refs.[1]-[3] a classical behaviour of quantum systems of the type (11)
(in particular the decoherence) is explained by means of the Feynman-Vernon
influence functional. This functional results from an average over 
environmental
degrees of freedom. We could derive it explicitely from the solution 
of eq.(20).
There is another theory of decoherence.The authors
\cite{Ghi2}\cite{Gisi} explain the emergence of classical
properties through an addition of some stochastic terms to the
Schr\"odinger equation. In spite of the ardent opposition of the
two groups ( see the letters in April 1993 issue of Physics Today \cite{Lett})
we would like to reconcile the two approaches.We restrict ourselves here to
the model (11). In eq.(20) the time evolution of the coordinate
$Q$ is influenced by the environment through the additional noise terms
$N_{R}$ and $N_{D}$.
$N_{R}$ (eq.(16)) can be considered as a
randomness resulting from the Feynman's sum over paths \cite{hab3} (
weighted sum over paths instead
of the classical extremal path). The "deterministic"
noise $N_{D}$ ( eq.(15) ) is produced by the environment of oscillators
evolving in a deterministic way according to the Schr\"odinger evolution
determined by the Hamiltonian (11).
We can show that similarly as in the model
of Ford et al \cite{Kac} the environment can
influence the system as a random force of the form of the white noise.
 We note first that if in eq.(18) $v_{k}\approx $const and
 $\omega_{k}\sim k$ then the noise $N_{D}(s)$ is strongly correlated only at
 coinciding arguments. Under these assumptions on $v_{k}$ and $\omega_{k}$
the dissipation kernel $ K(s)$
  in eq.(19)  can now be approximated by (such a behaviour of $K$ is
  sometimes called the Ohm dissipation)

\begin{displaymath}
K(s)=a\delta(s)
\end{displaymath}
where $a$ is a positive constant.In fact, we get such a formula  if
$\omega_{k}=k\omega$,where $k\in[0,\infty]$, and $v_{k}^{2}=\frac{a}{\pi}
\omega \triangle k$. In such a case the sum over infinitesimal
$\triangle k$ can be replaced by an integral
\begin{displaymath}
\int_{0}^{\infty}dkexp(-ik\omega s)=\frac{\pi}{\omega}\delta(s)
-\frac{i}{s\omega}
\end{displaymath}

Then, the term $s^{-1}$ is considered as small in comparison to $\delta(s)$.
The corresponding choice of $v_{k}^{2}$ leads to the following
approximation to $N_{D}$ (again the term $(t-s)^{-1}$ is neglected)
\begin{equation}
E[N_{D}(s)N_{D}(t)]=0
\end{equation}
\begin{displaymath}
E[\overline{N_{D}(s)}N_{D}(t)]=\hbar\epsilon \delta(t-s)
\end{displaymath}
where
\begin{displaymath}
\epsilon=\frac{a}{2 \omega_{0}}
\end{displaymath}
So,$N_{D}$ is the white noise. We denote the
corresponding Brownian motion by $B_{D}$ i.e.
\begin{displaymath}
\frac{dB_{D}}{ds}=N_{D}
\end{displaymath}
Next, let us consider the "quantum" noise $N_{R}$ (eq.(16)) in more 
detail. Let us
compute its correlation functions
\begin{equation}
E[N_{R}(s)N_{R}(\tau)]=\frac{\hbar}{2\omega_{0}}\sum_{k}v_{k}^{2}
(exp(-i\omega_{k}|\tau -s|) - exp(-i\omega_{k}(\tau+s)) )
\end{equation}
\begin{displaymath}
E[\overline{N_{R}(s)}N_{R}(\tau)]=\hbar min(s,\tau)\frac{1}{\omega_{0}}
\sum_{k}\omega_{k}v_{k}^{2}exp(i\omega_{k}(s-\tau))
\end{displaymath}
In the approximation of the Ohm dissipation we get by the same 
arguments as applied
at the derivation of eq.(21)
\begin{equation}
E[N_{R}(s)N_{R}(\tau)]=\hbar \epsilon \delta(s-\tau)
\end{equation}
 Eq.(23) means that $N_{R}$ is the real white noise (the second of the
 expectation values in eq.(22) does not enter any subsequent formulas).
We denote the integral of $N_{R}$ by $B_{R}$ (the real Brownian motion 
independent
of $B_{D}$).
Now,eq.(20) reads
\begin{equation}
dQ=-i\omega_{0}Qds-aQds-dB_{D}(s)- dB_{R}(s)+\lambda\sigma db(s)
\end{equation}
( all the noise terms on the right hand site of eq.(24) are independent).

The dynamics of a subsystem is usually investigated in terms of the time
evolution of its density matrix. We define the density matrix of a subsystem
( with coordinates X ) as
\begin{equation}
\rho_{P}(X,X^{\prime})=\int\prod_{k}dx_{k}\rho(X,x;x,X^{\prime})
\end{equation}
We make the usual assumption that initially the system and the environment
are separated i.e. $\rho$ is of the product form
\begin{displaymath}
\rho(X,x;x,X^{\prime})=exp(-\frac{\omega_{0}}{2\hbar}(X^{2}+X^{\prime 2}))
\overline{\nu(X)}\nu(X^{\prime})\rho_{0}(x,x)
\end{displaymath}
We have expressed the initial wave function of the subsystem as a product
of the ground state wave function and arbitrary $\nu(X)$ (  this is a proper
choice for the class of models (10)-(11)). We choose for $\rho_{0}$ the
probability density in the ground state (7). This is a reasonable assumption
for the reservoir at zero temperature ( the stochastic formalism
of sec.2 at positive temperature is discussed in \cite{Hab4}).

We can express in a simple way the evolution of subsystem$^{\prime}$s density
matrix in terms of the Brownian motion
\begin{equation}
\begin{array}{l}
\rho_{P}(t;X,X^{\prime})=
\cr
\int\prod_{k} dx_{k}
exp(-\frac{\omega_{0}}{\hbar}(X^{2}+X^{\prime 2}))|\chi_{0}(x)|^{2}
\overline{E[\nu(Q_{t}(X))]}E[\nu(Q_{t}(X^{\prime}))]
\end{array}
\end{equation}
For the model (11) the stochastic process $Q$ is the solution of the equation
(20).
When we apply the definition of the noise $N_{D}$ (eqs.(15) and (17)) then
the integration over $x_{k}$ can be expressed as an average over $N_{D}$
\begin{equation}
\rho_{P}(t;X,X^{\prime})
\simeq exp(-\frac{\omega_{0}}{2 \hbar}(X^{2}+X^{\prime 2}))
E_{D}[\overline{E[\nu(Q_{t}(X))]}E[\nu(Q_{t}(X^{\prime}))]]
\end{equation}
where E[..] means the expectation value with respect to the Brownian motion $b$
and $E_{D}$ means that the integration over x is replaced by an expectation
value with respect to the complex white noise $N_{D}$. An application of the
standard Ito stochastic calculus ( see e.g.\cite{Gikh})
 shows that $\nu(Q_{t})$ fulfills the Ito equation
\begin{equation}
\begin{array}{l}
d \nu=\frac{\partial \nu}{\partial Q}(-i\omega_{0}Qds -aQds-dB_{D}(s)
-dB_{R}+\lambda\sigma db)+
\cr
+\epsilon\frac{\hbar}{2} \frac{\partial^{2}\nu}{\partial Q^{2}}(1+i)ds
\end{array}
\end{equation}
The wave function $\psi(X)=exp(-\frac{\omega_{0}X^{2}}{2 \hbar})\nu(X)$
satisfies the stochastic Schr\"odinger equation
\begin{equation}
d\psi=-\frac{i}{\hbar}H_{0}\psi ds-\frac{\epsilon}{\hbar}H_{0}\psi ds
 -L\psi( dB_{D}(s)+dB_{R}(s))+\lambda\sigma L \psi db(s)
\end{equation}
where $H_{0}$ is the Hamiltonian of the oscillator (the formula (6) for the
$\omega_{0}$- oscillator)
\begin{displaymath}
L=\frac{\partial}{\partial X}+\frac{\omega_{0}}{\hbar}X
\end{displaymath}
We can derive the master equation for $\rho_{P}$ from eq.(27) without any
further approximation. For this purpose we differentiate $ \rho_{P} $ 
in eq.(27)
and apply eq.(29). Then, elementary rules of the
stochastic calculus \cite{Gikh} lead to the formula
\begin{equation}
\partial_{t}\rho_{P}=-\frac{i}{\hbar}[H_{0},\rho_{P}]-\frac{\epsilon}{\hbar}
(H_{0} \rho_{P} +\rho_{P} H_{0})
+ \hbar\epsilon L \rho_{P} L^{+}
\end{equation}
We have got the master equation exactly in the Lindblad form \cite{Lind}
as
\begin {displaymath}
H_{0}=\frac{\hbar^{2}}{2}L^{+}L
\end{displaymath}
The approximation which leads to the stochastic equation (29) is
a Markovian approximation implying linear differential equations for
the density matrix.The Lindblad form of the evolution equation ensures that
the evolution preserves the normalization of the density matrix and its
positivity.

We consider eq.(30) in the energy representation. Let $|k>$ be the
k-th eigenstate of $H_{0}$. Then, eq.(30) for matrix elements reads
(for a derivation it is sufficient to notice that L is proportional to
the anihilation operator)
\begin{equation}
\begin{array}{l}
\partial_{t}<j|\rho_{P}|k>=-i\omega_{0}(j-k)<j|\rho_{P}|k> -
\epsilon \omega_{0}(j+k)<j|\rho_{P}|k>+
\cr
+ 2\epsilon\omega_{0}\sqrt{(j+1)(k+1)}<j+1|\rho_{P}|k+1>
\end{array}
\end{equation}
For large j and k and a wide class of operators $\rho_{P}$ (in particular
for operators whose matrix elements have a limit $\hbar \rightarrow 0$)
$<j|\rho_{P}|k>\simeq<j+1|\rho_{P}|k+1>$. Applying this approximation
to the last term in eq.(31) we obtain the solution
\begin{equation}
<j|\rho_{P}(t)|k>\simeq exp(-\frac{i}{\hbar}(E_{j}-E_{k})t-
\frac{\epsilon}{\hbar}(\sqrt{E_{j}}-\sqrt{E_{k}} )^{2}t)<j|\rho_{P}(0)|k>
\end{equation}
here we denoted by $E_{k}$ the k-th energy eigenvalue.
Eq.(32) shows that for macroscopic energies the density
matrix tends exponentially fast (with the increase of time) to its diagonal.
So,if we take in eq.(26) $\nu=\sum _{p=P}^{\infty}c_{p}{\cal{H}}_{p}$,
(where $|p>={\cal{H}}_{p}$ is the Hermite polynomial of order p) then for large
P and large t the density matrix tends to 
$\sum_{p=P}^{\infty}|c_{p}|^{2}|p><p|$.
We can see that the interference disappears on a time scale proportional
to the inverse of the energy difference of the eigenstates in the sum $\nu$.
Hence, there is no interference of states with macroscopically
distinguishable energies.
There is still another interpretation of the result.Assume that in
 the framework of quantum mechanics of closed systems
with the conventional wave packet reduction postulate an apparatus
performed a measurement of the energy eigenvalues,but the results
are unknown. Then ,the state of the system after the measurement
is described by the solution (32) at large time. In this sense the
interaction with an environment produces the same effect as a
measurement.

If we consider eq.(30) in the momentum representation and assume
 $\omega_{0}X^{2}\simeq 0$ ,so that we can neglect the oscillator potential
energy in comparison with the kinetic energy,
then from eq.(30) we get approximately
\begin{equation}
\partial_{t}<p|\rho_{P}|p^{\prime}>\simeq
-\frac{\epsilon}{2\hbar}(p-p^{\prime})^{2}<p|\rho_{P}|p^{\prime}>
\end{equation}
Eq.(33) means a localization in momentum.If on the other hand we
assume that $\omega_{0}X^{2}$ is large in comparison with the
kinetic energy, then neglecting the momenta in eq.(30) we get in
the position representation
\begin{equation}
\partial_{t}<X|\rho_{P}|X^{\prime}>\simeq-\frac{\epsilon\omega_{0}^{2}}{\hbar}
(X-X^{\prime})^{2}<X|\rho_{P}|X^{\prime}>
\end{equation}
i.e. a decoherence in space.

\section{Summary and discussion}
We have discussed a description  of  the dynamics of
a subsystem which clearly displays the dissipation and noise in the
subsystem.
 It is shown in a simple model  that some classical properties can
 emerge in
 quantum systems from an interaction with a large (quantum) environment.
  In particular, we have derived  a stochastic Schr\"odinger
equation discussed in the theory of measurement.

  The theory of an interaction of microscopic and macroscpic bodies
 based on the stochastic Schr\"odinger equation
 has some appealing features. It describes the dynamics of decoherence
 and localization in a model-independent way.However, we think that the
 microworld and macroworld will meet in the conventional quantum mechanics.
 In such a case the problem of the  consistency of the deterministic
 ( in the sense of the Schr\"odinger time evolution) and stochastic
 evolutions
 will emerge. This was our motivation for an investigation of the stochastic
 Schr\"odinger equation as a consequence of a deterministic model. 
 The stochastic
 Schr\"odinger equation derived in sec.4 seems to be a realistic approximation
 to the time evolution in a reservoir. Surprisingly, it resembles a model
  of ref.
 \cite{Ghi2} for a continuous spontaneous localization during the energy
  measurement. A measurement of another observable needs another
  environment which in general would be difficult to define explicitly.
  However, it is important to enquire whether we can reduce the number
  of basic principles of quantum mechanics without any modification of
  the Schr\"odinger equation and the superposition principle.

\end{document}